\documentclass[12pt]{article}
\usepackage{latexsym}
\usepackage{amsfonts}
\usepackage{amsmath}
\usepackage{amsmath,amssymb,calc,amsfonts}
\usepackage{latexsym}
\usepackage{graphicx,calc,epsfig}

\begin{document}

\title{NSF 2.0: A spin weight zero formulation of General Relativity}
\author{Melina Bordcoch$^1$ \, Carlos Kozameh$^2$ \, Alejandra Rojas$^1$ \\
$^{1}${\small {FaCEN, Universidad Nacional de Catamarca, 4700 Catamarca,
Argentina} }\\
$^{2}${\small {FaMAF, Universidad Nacional de C\'{o}rdoba, 5000 C\'{o}rdoba,
Argentina}}}
\date{}
\maketitle

\begin{abstract}
We present a set of three PDEs for three real scalars that are equivalent to
the full Einstein equations without any symmetry assumptions. The main
variables in this formulation are null surfaces and a conformal factor.

Furthermore, for asymptotically flat spacetimes the free data (representing
gravitational radiation) enters as the source term in the resulting
equations. This could be important for an asymptotic quantization procedure.
\end{abstract}

\section{Introduction}

The Null-Surface Formulation of general relativity was developed in the 80's%
\cite{0a} and 90's\cite{0b,0c} with a double purpose. On one hand it
provided a generalization of self-dual General Relativity (GR) \cite{1} to
full GR. It also presented GR not as a field theory but as a theory of
surfaces on a four-manifold coupled to a scalar function. These surfaces
were characteristic surfaces of some conformal metric. The scalar function
played the role of a conformal factor converting the conformal metric into a
vacuum metric (with a straightforward generalization if matter were present).

The resulting field equations for NSF were given on the bundle of null
directions, with the spacetime as the base space and the fiber given by the
sphere of null directions. Furthermore, introducing the concept of spin
weight in the fiber space the resulting equations were labeled according to
this number. We recall that a sw-0 object is a scalar under rotations on the
sphere, a sw-1 object represents a vector, etc. The original equations were
then given as one sw-0 equation, two sw-1 equations and and two sw-3
equations. This set of equations were equivalent to the Einstein equations
albeit in a completely different form and for completely different variables.

The purpose of this note is to show that one can reformulate NSF as a set of
three sw-0 equations. It is remarkable that full GR without any symmetry
assumptions can be written as three PDEs for three real functions on the
bundle of null directions.

In Section II we first give a brief review of the NSF including some results
that differ the original derivations. However, most of the details are left
for the references contained therein. We then introduce the new set of
equations equivalent to the old ones. The details of the calculations are
given in the appendix. In Section III we present peeling in NSF, i.e., the
behavior of our variables as one approaches null infinity and show how the
free data representing the gravitational radiation enters as the source term
in the equations. We also show that the field equations adopt an extremely
simple form if we assume that the spacetime points are close to null
infinity and we give a perturbation procedure to obtain the null surfaces
good up to second order. Finally we close this work with some comments on
possible use of these results.

\section{A revised NSF.}

The null-surface approach to general relativity is a formulation based on
the introduction of two real functions, $Z$ and $\Omega$, living on the
bundle of null directions over $M$; i.e., on $M\times S^{2}$ with $x^{a}$ in
$M$, and $\left( \zeta,\bar{\zeta}\right) $ on $S^{2}$. These two functions
capture the information contained in an Einstein metric: $Z(x^{a},\zeta,\bar{%
\zeta})$ encodes the conformal structure of the Einstein space-time and
singles out a preferred member of the conformal class of metrics, while $%
\Omega(x^{a},\zeta,\overline{\zeta})$ represents the appropriate conformal
factor that turns the preferred member into an Einstein metric. In the
following, we show how these two functions are introduced. First, we present
the original version and then the spin weight zero version of the equations,
showing how the number of expressions is diminished from four complex
equations to two real ones.

\subsection{The Kinematical Structure}

On a manifold $M$ we consider an $S^{2}$-family of functions $Z(x^{a},\zeta,%
\overline{\zeta})$ and its level surfaces $Z=const.$ such that the equation
\begin{equation}
g^{ab}(x^{d})\partial_{a}Z(x^{d},\zeta,\bar{\zeta})\partial_{b}Z(x^{d},\zeta,%
\bar{\zeta})=0  \label{1}
\end{equation}
can be solved for a metric $g^{ab}(x^{a})$ for all values of $\left( \zeta,%
\overline{\zeta}\right) $. The surfaces $Z=const.$ are then null surfaces of the metric $g^{ab}$, i.e., for every fixed
value of $\left( \zeta,\overline{\zeta}\right) $, the equation $%
Z(x^{a},\zeta,\overline{\zeta})=u$ represents a null foliation of the
spacetime $(M,g^{ab})$. It is clear that kinematical conditions must be
imposed on $Z$ if we want a solution $g^{ab}$ of Eq.~(\ref{1}). To obtain
the solution as well as the conditions we consider eq. (\ref{1}) and small
deviations $(\zeta+d\zeta,\overline{\zeta}+d\overline{\zeta})$,of the
equation, i.e., we take derivatives of (\ref{1}) with respect to $\left(
\zeta,\overline{\zeta}\right) $. By repeated differentiation with respect to
$\zeta$ and $\bar{\zeta}$, and using the independence of the metric on the
variables $\left( \zeta,\overline{\zeta}\right) $, we obtain our two main
results: the explicit metric components and the metricity conditions imposed
on $Z$ for a solution to Eq.(\ref{1}) to exist.

The metric components are obtained on a natural class of coordinate systems
(one for each value of $\left( \zeta,\overline{\zeta}\right) $). This
special class of null coordinates $\theta^{i}(x^{a},\zeta,\overline{\zeta})$%
, $i=0,+,-,1$, is defined by the equation
\begin{equation}
\theta^{i}(x^{a},\zeta,\overline{\zeta})=(Z,\eth Z,\bar{\eth}Z,\bar{\eth}%
\eth Z).  \label{2}
\end{equation}
In the above equation, the operators $\eth$ and $\overline{\eth}$ are the $%
\zeta,\overline{\zeta}$ covariant derivatives on the sphere holding constant
$x^{a}$~\cite{3}. Eq.~(\ref{2}) should be interpreted as a coordinate
transformation $x^{a}\rightarrow\theta^{i}$ for every fixed value of $\left(
\zeta,\overline{\zeta}\right) $. In the following we will use indices $a$
for the $\left( \zeta,\overline{\zeta}\right) $ independent coordinates $%
x^{a}$ and $i$ for the $\left( x^{a},\zeta,\overline{\zeta}\right) $
dependent $\theta^{i}$. The inverse transformation is then given by%
\begin{equation*}
x^{a}=x^{a}\left( \theta^{i},\zeta,\overline{\zeta}\right) .
\end{equation*}

Using the gradient basis $\partial_{a}\theta^{i}$, together with the dual
basis $e_{i}^{a}=\partial_{i}x^{a}$ the metric $g^{ab}$ can be expressed in
the new coordinates as $g^{ab}(x)=g^{ij}e_{i}^{a}e_{j}^{b}$ where the metric
components $g^{ij}$ are explicitly given below.

The $g^{ij}$ are all expressible in terms of two quantities, $\eth^{2}Z$ and
$g^{01}$. By assumption $g^{01}$ is a positive definite scalar and can be
written as $g^{01}=\Omega^{2}$. However, in the derivations it is more
useful to keep its original definition,%
\begin{equation}
g^{01}=g^{ab}(x^{a})\partial_{a}Z\partial_{b}\bar{\eth}\eth Z.  \label{3}
\end{equation}

The function $\eth^{2}Z$ defines the complex scalar $\Lambda$ via%
\begin{equation*}
\Lambda\left( \theta^{i},\zeta,\overline{\zeta}\right)
=\eth^{2}Z(x^{a}\left( \theta^{i},\zeta,\overline{\zeta}\right) ,\zeta,%
\overline {\zeta}).
\end{equation*}

As we will see below the metric components depend on $\partial_{i}\Lambda$
and $\eth$ and $\overline{\eth}$ derivatives .

The same operator $\eth$ can be written in the $\theta^{i}$ coordinates in
the following manner%
\begin{equation}
\eth=\eth^{\prime}+\eth\theta^{i}\partial_{i}  \label{5}
\end{equation}
where $\eth^{\prime}$ is the usual $\zeta$ derivative holding $\theta^{i}$
fixed and $\eth\theta^{i}$ is given by
\begin{align}
\eth\theta^{0} & =\theta^{+}  \notag \\
\eth\theta^{+} & =\Lambda  \label{6} \\
\eth\theta^{-} & =\theta^{1}  \notag \\
\eth\theta^{1} & =\overline{\eth}\Lambda-2\theta^{+}.  \notag
\end{align}

Explicitly we have $g^{ij}=g^{01}h^{ij}[\Lambda]$ with $h^{ij}$ given by
\begin{equation*}
h^{00}=h^{0+}=h^{0-}=0
\end{equation*}%
\begin{equation*}
h^{01}=-h^{+-}=1
\end{equation*}%
\begin{equation*}
h^{++}=\bar{h}^{--}=-\partial_{1}\Lambda
\end{equation*}%
\begin{equation}
h^{+1}=\bar{h}^{-1}=\frac{1}{2}\frac{A-\frac{1}{2}\overline{A}\partial
_{1}\Lambda}{\left( 1-\frac{1}{4}\partial_{1}\Lambda\partial_{1}\overline{%
\Lambda}\right) }  \label{7}
\end{equation}%
\begin{equation*}
h^{11}=-\frac{\left[ \frac{1}{2}\left( \partial_{1}\overline{\eth}%
^{2}\Lambda+(h^{ij}-h^{11}\delta_{1}^{i}\delta_{1}^{j})\partial_{i}\Lambda%
\partial_{j}\overline{\Lambda}\right) +h^{-i}\partial_{i}\overline{\eth}%
\Lambda+h^{+i}\partial_{i}\eth\overline{\Lambda}+2\right] }{\left( 1+\frac{1%
}{2}\partial_{1}\Lambda\partial_{1}\overline{\Lambda }\right) },
\end{equation*}

where

\begin{equation*}
A=-\partial_{1}\overline{\eth}\Lambda+\partial_{+}\Lambda-h^{--}\partial
_{-}\Lambda.
\end{equation*}
The metric can then be written as%
\begin{equation}
g^{ab}(x^{a})=g^{01}h^{ij}e_{i}^{a}e_{j}^{b}.  \label{8}
\end{equation}

\subsection{The metricity conditions.}

The scalars $g^{01}$ and $Z$ are not arbitrary nor independent of each
other. They must satisfy two coupled differential equations, referred to as
the metricity conditions. The first one is obtained applying the $\eth$
operator to $g^{01}$, i. e., to the expression (\ref{3})
\begin{equation}
\eth g^{01}=g^{ab}\left( \partial_{a}\theta^{+}\partial_{b}\theta
^{1}+\partial_{a}\theta^{0}\partial_{b}\overline{\eth}\Lambda\right)
\label{9}
\end{equation}
which in components gives
\begin{equation}
\eth g^{01}=g^{01}\left( \partial_{1}\overline{\eth}\Lambda+h^{+1}\right) .
\label{10}
\end{equation}
Inserting explicitly the value of $h^{+1}$ in Eq.(\ref{10}) gives
\begin{equation}
\partial_{1}\overline{\eth}\Lambda+\partial_{+}\Lambda=M_{I}  \label{a}
\end{equation}
with
\begin{equation*}
M_{I}=2\left( 1-\frac{1}{4}\partial_{1}\Lambda\partial_{1}\overline{\Lambda }%
\right) \eth\ln g^{01}-\partial_{1}\bar{\Lambda}\partial_{-}\Lambda-\frac {1%
}{2}(\partial_{1}\eth\bar{\Lambda}-\partial_{-}\bar{\Lambda}-\partial
_{1}\Lambda\partial_{+}\bar{\Lambda}-\partial_{1}\overline{\Lambda}%
\partial_{1}\overline{\eth}\Lambda)\partial_{1}\Lambda.
\end{equation*}

For the second one we apply the $\eth^{3}$ operator to the expression (\ref%
{1}) obtaining,
\begin{equation}
g^{ab}(3\partial_{a}\theta^{+}\partial_{b}\Lambda+\partial_{a}\theta
^{0}\partial_{b}\eth\Lambda)=0  \label{11}
\end{equation}
which in components gives

\begin{equation}
\partial_{1}\eth\Lambda-3\partial_{-}\Lambda=M_{II}.  \label{12}
\end{equation}
with 
$$M_{II}=3\partial_{1}\Lambda\left( \partial_{+}\Lambda-h^{+1}\right) .$$

The two metricity conditions constitute the requirement on $Z$ in order for
a metric $g^{ab}$ to exist and not depend on $\left( \zeta,\overline{\zeta }%
\right) $ and such that $Z=const$ are characteristic surfaces of the metric,
for every value of $\left( \zeta,\overline{\zeta}\right) $. They leave $%
g^{01}$ undetermined up to a factor dependent only on $x^{a}$. The second
condition has another geometrical meaning and it is related to seminal work
done by E.\ Cartan on solutions of certain PDEs. To briefly review this
second point of view we start our construction from the equation%
\begin{equation}
\eth^{2}Z=\Lambda(Z,\eth Z,\bar{\eth}Z,\bar{\eth}\eth Z,\zeta,\overline{%
\zeta })  \label{4}
\end{equation}
where $\Lambda$ is assumed to be a given function of the six coordinates $%
\left( \theta^{i},\zeta,\overline{\zeta}\right) $ which satisfies the
integrability condition%
\begin{equation}
\overline{\eth}^{2}\Lambda=\eth^{2}\overline{\Lambda},  \label{I}
\end{equation}
Note that the coordinates $x^{a}$ have disappeared. If we analyze the
solution space of this equation we find that there are four constants of
integration and the solution can be written as

\begin{equation}
u=Z\left( x^{a},\zeta,\overline{\zeta}\right)
\end{equation}
(the spacetime points reenter as the solution space of this master
equation). Furthermore, if we want to characterize the special class of
functions that are equivalent to each other via diffeomorphisms $%
u^{\prime}=F\left( u,\zeta,\overline{\zeta}\right) $ and $%
\zeta^{\prime}=G\left( \zeta ,\overline{\zeta}\right) $ (called fiber
preserving transformations) we find that they should give the same conformal
structure on the solution space, i.e., they should give the same conformal
metric defined above. The last condition (\ref{12}) is then the requirement
that the normal metric connection has vanishing torsion \cite{4}.

From our point of view, this fiber preserving transformation $Z^{\prime
}=F\left( Z,\zeta,\overline{\zeta}\right) $ and $\zeta^{\prime}=G\left(
\zeta,\overline{\zeta}\right) $ clearly gives the same conformal structure
since for a fixed value of $\left( \zeta,\overline{\zeta}\right) $, $%
Z=const. $ and $Z^{\prime}=const.$ are both null surfaces for the same
metric.

\subsection{The real (sw-0) metricity conditions.}

We note that equation (\ref{10}) is a complex pde for a single real variable
$g^{01}$. In principle there could be no solutions unless the integrability
conditions are identically satisfied. This is indeed the case. Taking $%
\overline{\eth}$ of Eq. (\ref{9}), we obtain:

\begin{equation}
2\bar{\eth}\eth g^{01}\ =g^{ab}(\partial_{a}\bar{\eth}^{2}\Lambda\partial
_{b}\theta^{0}-\partial_{a}\Lambda\partial_{b}\bar{\Lambda}),  \label{13}
\end{equation}
or%
\begin{equation}
\bar{\eth}\eth g^{01}\ =\frac{1}{2}g^{01}(\partial_{1}\bar{\eth}%
^{2}\Lambda-h^{ij}\partial_{i}\Lambda\partial_{j}\bar{\Lambda}).  \label{14}
\end{equation}
Thus, $\bar{\eth}\eth g^{01}=\eth\bar{\eth}g^{01}$ if $\bar{\eth}%
^{2}\Lambda=\eth^{2}\bar{\Lambda}$, but this is true from the starting
assumption, equation (\ref{I}).

Note also that we could \ use the real Eq.(\ref{14}) instead of (\ref{10})
if we are interested in regular solutions for $g^{01}$. This follows from
the fact that the only regular solution of $\overline{\eth}F^{1}=0$ for a
s.w.1 function $F^{1}$is $F^{1}=0$. Thus, we could either use (\ref{10}) or (%
\ref{14}) as our first metricity condition.

With this in mind, we expect the same feature from our second metricity
condition, Eq. (\ref{12}). Directly from
\begin{equation*}
\bar{\eth}^{3}\eth^{3}g^{ab}(x^{d})\partial_{a}Z(x^{d},\zeta,\bar{\zeta }%
)\partial_{b}Z(x^{d},\zeta,\bar{\zeta})=0
\end{equation*}
(see the Appendix I for detailed calculations) we get
\begin{align}
& \partial_{1}\bar{\eth}\eth\bar{\eth}^{2}\Lambda+6\partial_{1}\bar{\eth}%
^{2}\Lambda+9h^{1i}\partial_{i}\bar{\eth}^{2}\Lambda+3h^{+j}\partial_{j}\bar{%
\eth}^{3}\Lambda+3h^{-k}\partial_{k}\eth^{3}\overline{\Lambda}+  \label{18}
\\
& +h^{ij}\left( 6\partial_{i}\Lambda\partial_{j}\bar{\Lambda}+9\partial
_{i}\eth\overline{\Lambda}\partial_{j}\bar{\eth}\Lambda+\partial_{i}%
\overline{\eth\Lambda}\partial_{j}\eth\Lambda+3\partial_{i}\eth\bar{\eth }%
\Lambda\partial_{j}\overline{\Lambda}+3\partial_{i}\bar{\eth}\eth \overline{%
\Lambda}\partial_{j}\Lambda\right) =0  \notag
\end{align}
\qquad\qquad\qquad$\qquad\qquad$

Again, and using the same argument as before, if we are interested in
regular solutions for \ $\Lambda$\ we can take Eq. (\ref{18}) instead Eq. (%
\ref{12}).

Summarizing, given two functions $\Lambda,$ $g^{01}$ that are solutions to (%
\ref{14}) and (\ref{18}), the level surfaces $Z\left( x^{a},\zeta ,\overline{%
\zeta}\right) =const.$ of the solution to (\ref{4}) are null hypersurfaces
of the metric defined by Eq. (\ref{1}). In the following sections of this
work we will use the spin weight zero real formulation, equations (\ref{14})
and (\ref{18}) to derive the field equations for $Z$.

\subsection{The field equations}

On this kinematical scheme we impose the (trace-free) vacuum Einstein
equations by $R_{ij}-\frac{1}{4}g_{ij}R=0$. These are a set of nine
equations for the metric components $g_{ij}$ in our coordinate system $%
\theta^{i}(x^{a},\zeta,\overline{\zeta})$ for arbitrary but fixed $(\zeta,%
\overline {\zeta})$. However, we can show that these equations are generated
by just one \ equation in the null bundle coordinates $(\theta^{i},\zeta,%
\overline{\zeta }).$ Consider the following contraction

\begin{equation}
R^{ab}(x^{a})\partial_{a}Z(x^{d},\zeta,\bar{\zeta})\partial_{b}Z(x^{d},\zeta,%
\bar{\zeta})=0.  \label{19}
\end{equation}

Taking $\eth$ and $\bar{\eth}$ on this equation we generate the remaining
eight trace-free vacuum Einstein equations in an analogous fashion as was
done on the previous subsection. Writing (\ref{19}) in terms of $\Lambda$
and $g^{01}=\Omega^{2}$ gives:
\begin{equation*}
\partial_{1}^{2}\Omega=\Omega R_{11}(\Lambda)
\end{equation*}
with $R_{11}(\Lambda)$, a component of the Ricci tensor of $h^{ij}$, given
by
\begin{align}
R_{11}(\Lambda) & =R^{ab}(\Lambda)\partial_{a}Z(x^{d},\zeta,\bar{\zeta }%
)\partial_{b}Z(x^{d},\zeta,\bar{\zeta})=\frac{1}{4q}\partial_{1}^{2}\Lambda%
\partial_{1}^{2}\bar{\Lambda}+\frac{3}{8q^{2}}(\partial_{1}q)^{2}-\frac{1}{4q%
}\partial_{1}^{2}q,  \label{20} \\
q & =1-\partial_{1}\Lambda\partial_{1}\bar{\Lambda}  \label{21}
\end{align}
The full set of Einstein equations can then be grouped together as
\begin{equation}
\partial_{1}^{2}\Omega-\Omega R_{11}(\Lambda)=0  \label{24}
\end{equation}
\begin{equation}
2\bar{\eth}\eth(\Omega^{2})-\Omega^{2}(\partial_{1}\bar{\eth}%
^{2}\Lambda-h^{ij}\partial_{i}\Lambda\partial_{j}\bar{\Lambda})=0  \label{22}
\end{equation}
\begin{align}
\partial_{1} & \bar{\eth}\eth\bar{\eth}^{2}\Lambda+6\partial_{1}\bar{\eth }%
^{2}\Lambda+9h^{1i}\partial_{i}\bar{\eth}^{2}\Lambda+3h^{+j}\partial_{j}\bar{%
\eth}^{3}\Lambda+3h^{-k}\partial_{k}\eth^{3}\overline{\Lambda}+  \label{23}
\\
& +h^{ij}\left( 6\partial_{i}\Lambda\partial_{j}\bar{\Lambda}+9\partial
_{i}\eth\overline{\Lambda}\partial_{j}\bar{\eth}\Lambda+\partial_{i}\bar{%
\eth }\overline{\Lambda}\partial_{j}\eth\Lambda+3\partial_{i}\eth\bar{\eth}%
\Lambda\partial_{j}\overline{\Lambda}+3\partial_{i}\bar{\eth}\eth \overline{%
\Lambda}\partial_{j}\Lambda\right) =0  \notag
\end{align}
The full system consists of two coupled PDEs for $\Omega$ and $\Lambda$ plus
the equation (\ref{23}) for $\Lambda$. From now on this set of equations
will be referred as the NSF equations.

Note that the first NSF equation, one real equation for $(\Omega,\Lambda)$, is equivalent to the
trace-free, Ricci flat equations, nine second order equations for six real
variables in the standard formulation (plus identities). However, the
metricity conditions must be incorporated in the NSF since they give the
conditions that the level surfaces $Z=const.$ are indeed null surfaces for
the metric of the spacetime. Thus, the three real equations of the NSF are
equivalent to the equations that yield Einstein spaces in GR.

\section{Asymptotic NSF}

Though any vacuum Einstein space-time can be investigated in this manner we
make the specialization, here, to asymptotically flat vacuum space-times. In
this case the geometrical meanings to the various quantities become more
focused and clearer and the differential equations become easier to handle.
We begin by assuming that the spacetime is asymptotically flat along null
directions with a null boundary $\mathcal{I}^{+}$. Null infinity can be
coordinatized with a Bondi coordinate system, $(u,\zeta,\bar{\zeta})$ with $%
u\in R$, the Bondi retarded time, and $(\zeta,\bar{\zeta})\in S^{2}$
labeling the null generators of $\mathcal{I}^{+}$. With this notation we can
give a precise meaning to the null surfaces described by $Z(x^{a},\zeta ,%
\overline{\zeta})=u=const$; they are taken to be the past null cones of the
points $(u,\zeta,\bar{\zeta})$ of $\mathcal{I}^{+}$. If in addition we set $%
\eth Z(x^{a},\zeta,\bar{\zeta})=\omega=const.$ we isolate a null geodesic in
this light cone. (Any null geodesic can be labelled by these five
parameters, the apex $(u,\zeta,\bar{\zeta})$ of the cone at $\mathcal{I}^{+}$%
, and its \textquotedblleft null angle", $(\omega,\bar{\omega})$.)
Furthermore, setting $\eth\bar{\eth}Z(x^{a},\zeta,\bar{\zeta})=R=const.$ we
describe a point on this null geodesic.

In addition to this meaning of $Z$, there is a dual meaning, namely, if the
space-time point $x^{a}$ is held constant but the $(\zeta,\bar{\zeta})$ is
varied over the $S^{2}$, we obtain a two-surface on $\mathcal{I}^{+}$, the
so-called light-cone cut of $\mathcal{I}^{+}$. It consists of all points of $%
\mathcal{I}^{+}$ reached by null-geodesics from $x^{a}$. $Z$ is then
referred to as the light-cone cut function. We now another interpretation,
not only of $Z(x^{a},\zeta,\overline{\zeta})$, but also of both, $%
\omega=\eth Z(x^{a},\zeta,\bar{\zeta})$ and $R=\eth\bar{\eth}Z(x^{a},\zeta,%
\bar{\zeta})$. $\omega$ is the \textquotedblleft stereographic angle" that
the light-cone cuts make with the Bondi $u=const$ cuts and $R$ is a measure
of the curvature of the cut.

\subsection{Peeling in NSF.}

Consider a Bondi slicing of $\mathcal{I}^{+}$ with the associated
coordinates $(u,\zeta,\bar{\zeta})$ and asymptotic shear $\sigma_{B}(u,\zeta,%
\bar{\zeta})$ for the associated Bondi null surfaces; then the asymptotic
shear $\sigma_{Z}$ of a light cone with apex at $x^{a}$ evaluated at $%
\mathcal{I}^{+}$ is given by
\begin{equation}
\sigma_{Z}(x^{a},\zeta,\bar{\zeta})=\sigma_{B}(Z,\zeta,\bar{\zeta})-\eth
^{2}Z\text{ },  \label{25}
\end{equation}
where $Z\left( x^{a},\zeta,\bar{\zeta}\right) $ is the light cone cut from $%
x^{a}$. This relationship is known as Sachs' theorem. From equations (\ref{4}%
) and (\ref{25}) we see that%
\begin{equation*}
\Lambda=\sigma_{B}(u,\zeta,\bar{\zeta})-\sigma_{Z}(u,\omega,\overline{\omega
},R,\zeta,\bar{\zeta}),
\end{equation*}
where we have used the coordinates $(u,\omega,\overline{\omega},R)$ to label
the apex $x^{a}$. We want to study the behavior of $\Lambda$\ as we move
the apex $x^{a}$ along a fixed null geodesic characterized by a given value
of $(u,\omega,\overline{\omega},\zeta,\bar{\zeta})$, i.e., along $R$. Since
our spacetime is asymptotically flat it is clear that as the apex $%
R\rightarrow \infty$ its light cone resembles more and more like a flat one
and thus it should have a vanishing shear in this limit. Thus%
\begin{equation*}
\lim_{R\rightarrow\infty}\Lambda=\sigma_{B}(u,\zeta,\bar{\zeta}).
\end{equation*}

Actually, this intuitive argument can be made precise if one uses the
optical equations for the shear and divergence of the light cone from $x^{a}$%
. One can show that (see Appendix 2) as $R\rightarrow\infty$ the leading
order term in $\sigma_{Z}$ is given by
\begin{equation}
\sigma_{Z}=\frac{\psi_{0}^{0}}{12R^{2}}+O\left( R^{-3}\right) ,  \label{S}
\end{equation}
and thus%
\begin{equation}
\Lambda=\sigma_{B}+\frac{\psi_{0}^{0}}{12R^{2}}+O\left( R^{-3}\right) .
\label{28}
\end{equation}
Remark: the fact that $\lim_{R\rightarrow\infty}\Lambda=\sigma_{B}(u,\zeta,%
\bar{\zeta})$ shows that $\Lambda$ explicitly depends on the free Bondi
data. Moreover, Eq.(\ref{28}) can be used to obtain all the metric
components and derived tensors in the asymptotic limit.

Other scalars have interesting asymptotic behavior as for example $%
\overline{\eth}\Lambda$ or $\overline{\eth}^{2}\Lambda$. To see this we
write the $\overline{\eth}$ operator in the $\theta^{i}$ coordinates as
\begin{equation}
\overline{\eth}=\overline{\eth}^{\prime}+\overline{\omega}\partial
_{u}+R\partial_{\omega}+\overline{\Lambda}\partial_{\overline{\omega}%
}+\left( \eth\overline{\Lambda}-2\overline{\omega}\right) \partial_{R}.
\label{c}
\end{equation}
Taking $\overline{\eth}$ of Eq. (\ref{28}) we get%
\begin{equation}
\overline{\eth}\Lambda=\overline{\eth}\sigma_{B}+O\left( R^{-1}\right)
\label{29}
\end{equation}

and $\overline{\eth}^{2}$(\ref{28}) gives:%
\begin{equation}
\overline{\eth}^{2}\Lambda=\overline{\eth}^{2}\sigma_{B}+O\left(
R^{0}\right) .  \label{30}
\end{equation}

We can make no asserts about this last decay just from kinematical
considerations. After we derive the field equations below we obtain the
explicit form of the second term on the r.h.s. of the equation.

\subsection{\protect\bigskip Field equations and free data.}

To derive the field equations we start with Eq.(\ref{22}), i.e.,
\begin{equation}
\partial_{R}\bar{\eth}^{2}\Lambda=2\frac{\bar{\eth}\eth\Omega^{2}}{\Omega^{2}%
}\ +h^{ab}\partial_{a}\Lambda\partial_{b}\bar{\Lambda}.  \label{32}
\end{equation}
Integrating (\ref{32}) over $R$ we obtain
\begin{equation}
\bar{\eth}^{2}\Lambda+\int_{R}^{\infty}[2\frac{\bar{\eth}\eth\Omega^{2}}{%
\Omega^{2}}\ +h^{ab}\partial_{a}\Lambda\partial_{b}\bar{\Lambda}]dR^{\prime
}=\left( \bar{\eth}^{2}\Lambda\right) _{R\rightarrow\infty},
\end{equation}
where ($\bar{\eth}^{2}\Lambda)_{R\rightarrow\infty}$ is a function that
depends on $(u,\omega,\overline{\omega},\zeta,\bar{\zeta})$. Using the
remaining field equations plus the peeling behavior of all the scalars
involved one can show that (see Appendix 3 for details)

\begin{equation}
\left( \partial_{u}\bar{\eth}^{2}\Lambda\right) _{R\rightarrow\infty }=%
\overline{\eth}^{2}\overset{\cdot}{\sigma}_{B}+\eth^{2}\overset{\cdot }{%
\overline{\sigma}}_{B}+\overset{\cdot}{\sigma}_{B}\overset{\cdot}{\overline{%
\sigma}}_{B}.  \label{limit}
\end{equation}
Using the fact that $\partial_{u}$ commutes with $\eth^{2}$ and $\overline {%
\eth}^{2}$ in the limit $R\rightarrow\infty$\ we finally obtain

\begin{equation}
\left( \bar{\eth}^{2}\Lambda\right) _{R\rightarrow\infty}=\overline{\eth }%
^{2}\sigma_{B}+\eth^{2}\overline{\sigma}_{B}+\int_{-\infty}^{u}\overset{\cdot%
}{\sigma}_{B}\overset{\cdot}{\overline{\sigma}}_{B}du.  \label{40}
\end{equation}
The constant of integration has been set equal to zero since we restrict the
BMS group to the Poincar\`{e} group by asking $\sigma\rightarrow0$ as $%
u\rightarrow-\infty$. The integral term $\int_{-\infty}^{\infty}\overset{%
\cdot}{\sigma}_{B}\overset{\cdot}{\overline{\sigma}}_{B}du$ is the total
energy radiated by the system and it is finite by assumption.\footnote{%
Note that there exists a difference in the plus sign of the las term of Eq. (%
\ref{40}) and the minus sign in the same term of the peeling of the main
equation (15) in \cite{5}}

The final equation reads
\begin{equation}
\bar{\eth}^{2}\Lambda+\int_{R}^{\infty}[2\frac{\bar{\eth}\eth\Omega^{2}}{%
\Omega^{2}}\ +h^{ij}\partial_{i}\Lambda\partial_{j}\bar{\Lambda}]dR^{\prime
}=\overline{\eth}^{2}\sigma_{B}+\eth^{2}\overline{\sigma}_{B}+\int_{-\infty
}^{u}\overset{\cdot}{\sigma}_{B}\overset{\cdot}{\overline{\sigma}}_{B}du
\label{41}
\end{equation}
where\ $\Omega$\ must be solved from the Einstein's equation (\ref{24}).\
Equation (\ref{41}) together with (\ref{23}) and (\ref{24}) is the full set
of NSF equations for asymptotically flat spacetimes. Note that Bondi shear $%
\sigma_{B}(u,\zeta,\bar{\zeta})$ enters as a source term in Eq.(\ref{41}).
This shows the explicit dependance of $\Lambda$ on the free data.

\subsection{Leading Order Equations}

We have shown above that $\lim_{R\rightarrow\infty}\Lambda=\sigma_{B}(u,%
\zeta,\bar{\zeta})$. We now study the behavior of $\bar{\eth}^{2}\Lambda$
in this limit. \ For this we study \ the integral term on the l.h.s. of Eq.(%
\ref{41}). From Eq. (\ref{20}) one can see that
\begin{equation*}
Q\left( \Lambda\right) =O\left( R^{-8}\right) +...
\end{equation*}
It then follows from Eq. (\ref{24}) that
\begin{equation}
\Omega=1+O\left( R^{-6}\right)  \label{31}
\end{equation}
One can also see that
\begin{equation}
h^{ij}\partial_{i}\Lambda\partial_{j}\bar{\Lambda}=O\left( R^{-3}\right)
\label{33}
\end{equation}
Thus, the leading order term for $\bar{\eth}^{2}\Lambda$ on the region $R>>1$%
, is given by%
\begin{equation}
\bar{\eth}^{2}\Lambda=\overline{\eth}^{2}\sigma_{B}+\eth^{2}\overline{\sigma
}_{B}+\int_{-\infty}^{u}\overset{\cdot}{\sigma}_{B}\overset{\cdot}{\overline{%
\sigma}}_{B}du+O\left( R^{-2}\right) .  \label{34}
\end{equation}
Note that if we keep the leading order term in the above equation and we
replace $u$ and $\Lambda$ for $Z$ and $\eth^{2}Z$ respectively we get a
field equation directly for $Z$, i.e.,
\begin{equation}
\bar{\eth}^{2}\eth^{2}Z=\overline{\eth}^{2}\sigma_{B}+\eth^{2}\overline {%
\sigma}_{B}+\int_{-\infty}^{Z}\overset{\cdot}{\sigma}_{B}\overset{\cdot }{%
\overline{\sigma}}_{B}du.  \label{35}
\end{equation}
Eq. (\ref{35}) is a real generalization of the good cut equation of $H$%
-space \cite{1}. It also resembles the nice section equation of Moreschi
\cite{6} and the Geroch-Winicour equation \cite{7}. Note however that the
above equation is different from the ones mentioned before. The final
conclusion is that the solution space of Eq. (\ref{35}) is different from
the ones obtained in the other formulations since there is a one to one
correspondence between points in the solution space and cuts of null
infinity. An analogous statement would be that the light cone cuts produced
by points of an asymptotically flat spacetime will satisfy Eq. (\ref{35}) in
the limit $R\rightarrow\infty$.

\subsection{A perturbative solution for $\Lambda.$}

Since the source term in Eq. (\ref{41}) is quadratic in the Bondi data $%
\sigma _{B}$ we want to obtain here an equation for which is correct up to
this second order. To do this we star with Eq. (\ref{24}) and solve
perturbately for $\Omega $, obtaining
\begin{equation*}
\Omega =1+\partial _{R^{\prime }}\Lambda \partial _{R^{\prime }}\overline{%
\Lambda }+\int\limits_{R^{\prime }}^{\infty }\int\limits_{R^{\prime \prime
}}^{\infty }\partial _{R^{\prime \prime \prime }}^{2}\Lambda \partial
_{R^{\prime \prime \prime }}^{2}\overline{\Lambda }dR^{\prime \prime \prime
}dR^{\prime \prime }+O(\Lambda ^{3}).
\end{equation*}

\bigskip Thus, the field equation for $\Lambda $ reads

\begin{equation}
\bar{\eth }^{2}\Lambda +\int\limits_{R}^{\infty }[\overline{\eth }\eth
\left( \partial _{R^{\prime }}\Lambda \partial _{R^{\prime }}\overline{%
\Lambda }+\int\limits_{R^{\prime }}^{\infty }\int\limits_{R^{\prime \prime
}}^{\infty }\partial _{R^{\prime \prime \prime }}^{2}\Lambda \partial
_{R^{\prime \prime \prime }}^{2}\overline{\Lambda }dR^{\prime \prime \prime
}dR^{\prime \prime }\right) +\eta ^{ij}\partial _{i}\Lambda \partial _{j}%
\bar{\Lambda}]dR^{\prime }=\bar{\eth }^{2}\sigma +\eth ^{2}\overline{\sigma }%
+\int\limits_{-\infty }^{u}\overset{\cdot }{\sigma }_{B}\overset{\cdot }{%
\overline{\sigma }}_{B}du  \label{42}
\end{equation}%
where $\eta ^{ij}$ represents the Minkowski spacetime metric. This equation
can be solved perturbately as follows. we first start with teh zero order
solution for $Z^{(0)}$,

\begin{align}
\sigma ^{(0)}& =0  \notag \\
\Lambda ^{(0)}& =0  \notag \\
\eth ^{2}Z^{(0)}& =0  \notag \\
Z^{(0)}& =x^{a}\ell _{a}  \label{43}
\end{align}%
The first order term $\Lambda ^{(1)}$ is then obtained from
\begin{align*}
\bar{\eth }^{2}\Lambda ^{(1)}& =\bar{\eth }^{2}\sigma (Z^{(0)},\zeta ,%
\overline{\zeta })+\eth ^{2}\overline{\sigma }(Z^{(0)},\zeta ,\overline{%
\zeta }) \\
\bar{\eth }^{2}[\Lambda ^{(1)}-\sigma (Z^{(0)},\zeta ,\overline{\zeta })]&
=\eth ^{2}\overline{\sigma }(Z^{(0)},\zeta ,\overline{\zeta })
\end{align*}%
and integrating at $x^{a}$ constant we get%
\begin{equation}
\Lambda ^{(1)}=\sigma _{B}(Z^{(0)},\zeta ,\overline{\zeta }%
)+\int\limits_{S^{2}}\overline{K}_{2,0}^{+}\eth ^{\prime 2}\bar{\sigma}%
(x^{a}l_{a}^{^{\prime }},\zeta ^{\prime },\overline{\zeta }^{\prime })dS^{2}
\label{44}
\end{equation}%
where $\overline{K}_{2,0}^{+}$ is the corresponding Green's function for the
$\bar{\eth }^{2}$ operator \cite{3}. Finally%
\begin{align}
\bar{\eth }^{2}\eth ^{2}Z^{(2)}& =\bar{\eth }^{2}\sigma (Z^{(1)},\zeta ,%
\overline{\zeta })+\eth ^{2}\overline{\sigma }(Z^{(1)},\zeta ,\overline{%
\zeta })+\int_{-\infty }^{Z_{0}}\overset{\cdot }{\sigma }(Z^{(0)},\zeta ,%
\overline{\zeta })\overset{\cdot }{\overline{\sigma }}(Z^{(0)},\zeta ,%
\overline{\zeta })du+  \notag \\
& +\int_{\infty }^{R}[\overline{\eth }\eth \left( \partial _{R}\Lambda
^{(1)}\partial _{R}\overline{\Lambda }^{(1)}\right) +\overline{\eth }\eth
\left( \int_{R^{\prime }}^{\infty }\int_{R^{\prime \prime }}^{\infty
}\partial _{R^{\prime \prime \prime }}^{2}\Lambda ^{(1)}\partial _{R^{\prime
\prime \prime }}^{2}\overline{\Lambda }^{(1)}dR^{\prime \prime \prime
}dR^{\prime \prime }\right)  \label{45} \\
& +\eta ^{ij}\partial _{i}\Lambda ^{(1)}\partial _{j}\overline{\Lambda }%
^{(1)}]dR^{\prime }  \notag
\end{align}

As we can see, the expression (\ref{45}) is a PDE on the sphere for the cut
function $Z^{(2)}$ is completely know from previous integrations. The
equation can be integrated using the corresponding Green's function for $%
\bar{\eth}^{2}\eth^{2}$.

\section{Summary and Conclusions}

In this work we have obtained a new derivation of the field equations for
NSF. By proper differentiation we obtain spin weight zero equations that are
equivalent to the original ones. The resulting equations constitute a set of
three real PDEs for three real functions on a six dimensional space. The
solution of those equations yield null surfaces on the spacetime,
(equivalent to the knowledge of a conformal metric) and the conformal factor
that yields a vacuum metric.

For asymptotically flat spacetimes one obtains equations where the source
term is the free Bondi data for gravitational radiation. This new result is
particularly important for several applications. Since the light cone cut $%
Z(x^{a},\zeta,\bar{\zeta})$ of any point $x^{a}$ of the spacetime must
satisfy this equation, one can define a family of light cone cuts associated
with a worldline by simply giving the parametric form of the worldline as $%
x^{a}(\tau)$ with $\tau$ the proper time of the worldline. In particular,
any candidate for a center of mass worldline should be a solution of Eq. (\ref%
{41}). In this context, none of the approaches that are present in the
literature satisfy this equation and it should be useful to incorporate Eq. (%
\ref{41}) in those approaches.

It is also worth pursuing further studies on the solution of Eq. (\ref{35}).
We have the machinery needed to investigate the dynamical properties of the
spacetime defined on the solution space of this equation.

Finally, we would like to discuss and issue that must be addressed each time
a new variable is introduced to describe General Relativity. What is the
advantage, if any, to present this non local variable instead of the metric
of the spacetime? After all, from knowledge of the metric one can construct
null geodesics, null cones, etc.

So far one can point out some advantages that come directly from the NSF
formalism; Since the equations yield global variables any perturbation (or
numerical) calculation of the solution modifies the null surfaces of the
spacetime at each order of the perturbation calculation, a task that is
virtually impossible with standard variables. Another advantage of the NSF\
is that the free data enters directly into the equations with a clear
physical meaning, namely, the data is the gravitational radiation that
reaches null infinity. One of the outstanding problems in a Cauchy
formulation of GR is to define what constitutes physically relevant data and
this issue is non existent in the NSF. At a classical level one can foresee
the role of NSF not as competing with the standard approach but rather as an
alternative formulation available to discuss global issues and/or solutions
without any symmetry.

The perception as to which is a good variable changes completely when
dealing with a quantum theory of gravity. In Loop Quantum Gravity the metric
field is not an observable since it is not a self adjoint operator. Instead,
non local variables such as holonomies around loops play \ a leading role in
the theory. Here NSF could play an important role for several reasons. By
construction the main variables are non local observables and the free data $%
\sigma_{B}$ is promoted to a quantum operator $\widehat{\sigma}_{B}$ that
obeys free commutation relations. Since $\lim_{R\rightarrow\infty}\Lambda=%
\sigma _{B}(u,\zeta,\bar{\zeta})$ one has in NSF a well defined variable
that asymptotically approaches the free quantum operator $\widehat{\sigma}%
_{B}$. This could be particularly useful for an S matrix approach to quantum
gravity.

\appendix
\vspace{1cm}
{\bf {\Large Appendix I}}
\vspace{.5cm}

In order to obtain equation (\ref{18}) we apply $\bar{\eth }^{3}\eth ^{3}$\
to $g^{ab}(x^{d})\partial _{a}Z\partial
_{b}Z=0$ which corresponds to applying $\bar{\eth }%
^{3}$ to (\ref{11})
\begin{equation*}
\bar{\eth }^{3}g^{ab}(3\partial _{a}\theta ^{+}\partial _{b}\Lambda
+\partial _{a}\theta ^{0}\partial _{b}\eth \Lambda )=0,
\end{equation*}%
obtaining,%
\begin{align}
0& =g^{ab}(\partial _{a}\bar{\eth }^{3}\eth \Lambda \partial _{b}u+3\partial
_{a}\bar{\eth }^{3}\Lambda \partial _{b}\omega +3\partial _{a}\bar{\eth }%
^{2}\eth \Lambda \partial _{b}\overline{\omega }+3\partial _{a}\bar{\eth }%
\eth \Lambda \partial _{b}\overline{\Lambda }+3\partial _{a}\bar{\eth }\eth
\overline{\Lambda }\partial _{b}\Lambda -  \notag \\
& -18\partial _{a}\bar{\eth }\Lambda \partial _{b}\overline{\omega }%
+9\partial _{a}\bar{\eth }^{2}\Lambda \partial _{b}R+\partial _{a}\bar{\eth }%
\overline{\Lambda }\partial _{b}\eth \Lambda -6\partial _{a}\Lambda \partial
_{b}\bar{\Lambda}+9\partial _{a}\eth \overline{\Lambda }\partial _{b}\bar{%
\eth }\Lambda ).  \label{15}
\end{align}%
Performing $\left[ \overline{\eth },\eth \right] F^{s}=2sF^{s}$ and
inserting $\bar{\eth }^{2}\Lambda =\eth ^{2}\overline{\Lambda }$, equation (%
\ref{15}) can be put in the form%
\begin{align}
0& =g^{ab}(\partial _{a}\bar{\eth }\eth \bar{\eth }^{2}\Lambda \partial
_{b}u+6\partial _{a}\bar{\eth }^{2}\Lambda \partial _{b}u+9\partial _{a}\bar{%
\eth }^{2}\Lambda \partial _{b}R+6\partial _{a}\Lambda \partial _{b}\bar{%
\Lambda}+9\partial _{a}\eth \overline{\Lambda }\partial _{b}\bar{\eth }%
\Lambda +  \notag \\
& +\partial _{a}\bar{\eth }\overline{\Lambda }\partial _{b}\eth \Lambda
+3\partial _{a}\bar{\eth }^{3}\Lambda \partial _{b}\omega +3\partial
_{a}\eth ^{3}\overline{\Lambda }\partial _{b}\overline{\omega }+3\partial
_{a}\eth \bar{\eth }\Lambda \partial _{b}\overline{\Lambda }+3\partial _{a}%
\bar{\eth }\eth \overline{\Lambda }\partial _{b}\Lambda )  \label{16}
\end{align}%
which is clearly a real equation. Setting (\ref{16}) in components we have,%
\begin{align*}
0& =\partial _{1}\bar{\eth }\eth \bar{\eth }^{2}\Lambda +6\partial _{1}\bar{%
\eth }^{2}\Lambda +9h^{1i}\partial _{i}\bar{\eth }^{2}\Lambda
+3h^{+j}\partial _{j}\bar{\eth }^{3}\Lambda +3h^{-k}\partial _{k}\eth ^{3}%
\overline{\Lambda }+ \\
& +h^{ij}\left( 6\partial _{i}\Lambda \partial _{j}\bar{\Lambda}+9\partial
_{i}\eth \overline{\Lambda }\partial _{j}\bar{\eth }\Lambda +\partial _{i}%
\bar{\eth }\overline{\Lambda }\partial _{j}\eth \Lambda +3\partial _{i}\eth
\bar{\eth }\Lambda \partial _{j}\overline{\Lambda }+3\partial _{i}\bar{\eth }%
\eth \overline{\Lambda }\partial _{j}\Lambda \right)
\end{align*}%
which is Eq. (\ref{18}) in the main text.

\vspace{1cm}
{\bf {\Large Appendix II}}
\vspace{.5cm}

In this Appendix we calculate the form of $\sigma_{Z}(u,\omega,\overline {%
\omega},R,\zeta,\overline{\zeta})$. We start from the equations for the
optical scalars associated with null geodesic congruences on a vacuum
spacetime%
\begin{equation}
\frac{\partial\rho}{\partial S}=\rho^{2}+\sigma_{Z}\overline{\sigma}_{Z}
\label{26}
\end{equation}%
\begin{equation}
\frac{\partial\sigma_{Z}}{\partial S}=2\rho\sigma_{Z}+\psi_{0}\text{ ,}
\label{27}
\end{equation}
with $S$ being the affine parameter and $\psi_{0}$ $\equiv$ $C_{1313}$ as
defined in the Newman-Penrose formalism \cite{8}.

In order to integrate the equations (\ref{26}) and (\ref{27}) we introduce a
perturbative scheme in terms of a strength factor, considered as $\epsilon
\ll1$, such that we can write
\begin{equation*}
\sigma=\sigma_{{}}^{(0)}+\epsilon\sigma_{{}}^{(1)}+\epsilon^{2}%
\sigma_{{}}^{(2)}+...
\end{equation*}%
\begin{equation*}
\rho=\rho^{\left( 0\right) }+\epsilon\rho^{\left( 1\right) }+\epsilon
^{2}\rho^{\left( 2\right) }+...
\end{equation*}
Then, we solve through a perturbation procedure.

Order zero: Solving equations (\ref{26}) and (\ref{27}) for Minkowski space $%
\sigma_{Z}^{(0)}=0$ and%
\begin{equation*}
\frac{\partial\rho^{(0)}}{\partial S}=(\rho^{\left( 0\right) })^{2}
\end{equation*}
whose solution is
\begin{equation*}
\rho^{\left( 0\right) }=-\frac{1}{S-S_{0}}
\end{equation*}
with $S_{0}$ representing the position of the apex of the light cone.

Order $\epsilon$: Writing equations (\ref{26}) and (\ref{27}) for $%
\rho^{\left( 1\right) }$ and $\sigma^{(1)}$ as
\begin{equation*}
\frac{\partial\rho^{(1)}}{\partial S}=2\rho^{\left( 0\right) }\rho^{\left(
1\right) }
\end{equation*}
\
\begin{equation*}
\frac{\partial\sigma^{(1)}}{\partial S}=2\rho^{\left( 0\right) }\sigma
^{(1)}+\frac{\psi_{0}^{0}}{S^{5}}
\end{equation*}
we find that the solutions are%
\begin{equation*}
\rho^{\left( 1\right) }=0
\end{equation*}%
\begin{equation*}
\sigma^{(1)}=\frac{\psi_{0}^{0}}{\left( S-S_{0}\right) ^{2}}\left[ -\frac{1}{%
2S^{2}}+\frac{2S_{0}}{3S^{3}}-\frac{S_{0}^{2}}{4S^{4}}+\frac {1}{12S_{0}^{2}}%
\right] \text{ }
\end{equation*}
Taking the limit $\lim_{S\rightarrow\infty}S^{2}\sigma^{(1)}$,
\begin{equation*}
\sigma_{Z}=\lim_{S\rightarrow\infty}S^{2}\sigma^{(1)}=\frac{\psi_{0}^{0}}{%
12S_{0}^{2}}.
\end{equation*}
The last step is to show that in the limit $\lim_{S\rightarrow\infty}\frac {%
dR}{dS}\rightarrow1$. This is true by virtue of the defining equation for $%
g^{01}$, namely%
\begin{equation*}
\frac{dR}{dS}=g^{01},
\end{equation*}
asymptotic simplicity implies $\lim_{S\rightarrow\infty}g^{01}\rightarrow1$.
Thus, the behavior of $\sigma_{Z}$ is $O(R^{-2})$, where $R$ is the
position of the apex. This last expression corresponds to equation (\ref{S})
of the main text.

\vspace{1cm}
{\bf {\Large Appendix III}}
\vspace{.5cm}

In this Appendix we obtain equation (\ref{limit}). Although the intervening
equations are long and the derivation cumbersome the only goal is to obtain
a limit using the peeling behavior of our variables. Most of the time this
peeling is easy to take into account but there is one term that, as we see
below, requires a bit of effort.

We first split the linear and upper order terms in $\Lambda $ in Eq. (\ref%
{23}) as,%
\begin{align}
& \partial _{1}\bar{\eth }\eth \bar{\eth }^{2}\Lambda +6\partial _{1}\bar{%
\eth }^{2}\Lambda +9\partial _{u}\bar{\eth }^{2}\Lambda -3\partial _{%
\overline{\omega }}\bar{\eth }^{3}\Lambda -3\partial _{\omega }\eth ^{3}%
\overline{\Lambda }+  \notag \\
& +9(h^{1+}\partial _{+}\bar{\eth }^{2}\Lambda +h^{1-}\partial _{-}\bar{\eth
}^{2}\Lambda +h^{11}\partial _{1}\bar{\eth }^{2}\Lambda )+  \notag \\
& +3(h^{++}\partial _{+}\bar{\eth }^{3}\Lambda +h^{+1}\partial _{1}\bar{\eth
}^{3}\Lambda +h^{--}\partial _{-}\eth ^{3}\overline{\Lambda }+h^{-1}\partial
_{1}\eth ^{3}\overline{\Lambda })+  \notag \\
& +h^{ab}\left( 6\partial _{a}\Lambda \partial _{b}\bar{\Lambda}+9\partial
_{a}\eth \overline{\Lambda }\partial _{b}\bar{\eth }\Lambda +\partial _{a}%
\bar{\eth }\overline{\Lambda }\partial _{b}\eth \Lambda +3\partial _{a}\eth
\bar{\eth }\Lambda \partial _{b}\overline{\Lambda }+3\partial _{a}\bar{\eth }%
\eth \overline{\Lambda }\partial _{b}\Lambda \right) =0.  \label{A3}
\end{align}

We use the commutation relations between $\partial _{i}$ and $\eth $ written
as%
\begin{equation*}
\lbrack \partial _{i},\eth ]=\delta _{i}^{+}(\partial _{0}-2\partial
_{1})+\delta _{i}^{1}\partial _{-}+\partial _{i}\Lambda \partial
_{+}+\partial _{i}\overline{\eth }\Lambda \partial _{1}\equiv \lbrack
\partial _{i},\eth ]^{0}+[\partial _{i},\eth ]^{1}
\end{equation*}%
and its complex conjugate, where $[\partial _{i},\eth ]^{0}\equiv \delta
_{i}^{+}(\partial _{0}-2\partial _{1})+\delta _{i}^{1}\partial _{-}$ and $%
[\partial _{i},\eth ]^{1}\equiv \partial _{i}\Lambda \partial _{+}+\partial
_{i}\overline{\eth }\Lambda \partial _{1}$ to reexpress the first term of (%
\ref{A3}), $\partial _{1}\bar{\eth }\eth \bar{\eth }^{2}\Lambda $ as
\begin{eqnarray*}
\partial _{1}\bar{\eth }\eth \bar{\eth }^{2}\Lambda &=&\bar{\eth }\eth
\partial _{1}\bar{\eth }^{2}\Lambda +\partial _{-}\bar{\eth }^{3}\Lambda
+\partial _{+}\eth \bar{\eth }^{2}\Lambda -\partial _{0}\bar{\eth }%
^{2}\Lambda +2\partial _{1}\bar{\eth }^{2}\Lambda + \\
&&+\left[ \partial _{1},\bar{\eth }\right] ^{1}\eth \bar{\eth }^{2}\Lambda +%
\bar{\eth }\left[ \partial _{1},\eth \right] ^{1}\bar{\eth }^{2}\Lambda +%
\left[ \overline{\eth },\partial _{-}\right] ^{1}\bar{\eth }^{2}\Lambda
\end{eqnarray*}%
and replacing in (\ref{A3})\ we get%
\begin{equation}
4\partial _{u}\bar{\eth }^{2}\Lambda -\partial _{\overline{\omega }}\bar{%
\eth }^{3}\Lambda -\partial _{\omega }\eth ^{3}\overline{\Lambda }+4\partial
_{R}\bar{\eth }^{2}\Lambda +\frac{1}{2}\bar{\eth }\eth \partial _{R}\bar{%
\eth }^{2}\Lambda +\frac{1}{2}M_{IIRe}=0  \label{mii}
\end{equation}%
with%
\begin{align}
M_{IIRe}& =9\left( h^{1+}\partial _{+}\bar{\eth }^{2}\Lambda
+h^{1-}\partial _{-}\bar{\eth }^{2}\Lambda +h^{11}\partial _{1}\bar{\eth }%
^{2}\Lambda \right) +  \notag \\
& +3\left( h^{++}\partial _{+}\bar{\eth }^{3}\Lambda +h^{+1}\partial _{1}%
\bar{\eth }^{3}\Lambda +h^{--}\partial _{-}\eth ^{3}\overline{\Lambda }%
+h^{-1}\partial _{1}\eth ^{3}\overline{\Lambda }\right) +  \notag \\
& +h^{ab}\left( 6\partial _{a}\Lambda \partial _{b}\bar{\Lambda}+9\partial
_{a}\eth \overline{\Lambda }\partial _{b}\bar{\eth }\Lambda +\partial _{a}%
\bar{\eth }\overline{\Lambda }\partial _{b}\eth \Lambda +3\partial _{a}\eth
\bar{\eth }\Lambda \partial _{b}\overline{\Lambda }+3\partial _{a}\bar{\eth }%
\eth \overline{\Lambda }\partial _{b}\Lambda \right) +C(\Lambda ).
\label{A4}
\end{align}%
where%
\begin{align}
C& =\left[ \partial _{1},\bar{\eth }\right] ^{1}\eth \bar{\eth }%
^{2}\Lambda +\bar{\eth }\left[ \partial _{1},\eth \right] ^{1}\bar{\eth }%
^{2}\Lambda +\left[ \overline{\eth },\partial _{-}\right] ^{1}\bar{\eth }%
^{2}\Lambda \\
& =\left[ \partial _{1},\bar{\eth }\right] ^{1}\eth ^{3}\Lambda +%
\left[ \partial _{1},\eth \right] ^{1}\bar{\eth }^{3}\Lambda +\left[
\overline{\eth },\partial _{-}\right] ^{1}\bar{\eth }^{2}\Lambda +\left[
\eth ,\partial _{+}\right] ^{1}\eth ^{2}\overline{\Lambda }+ \\
& +\left( \left[ \overline{\eth },\partial _{1}\right] ^{1}\Lambda \partial
_{+}+\left[ \overline{\eth },\partial _{1}\right] ^{1}\overline{\eth }%
\Lambda \partial _{1}+\partial _{1}\Lambda \left[ \overline{\eth },\partial
_{+}\right] ^{1}+\partial _{1}\overline{\eth }\Lambda \left[ \overline{\eth }%
,\partial _{1}\right] ^{1}\right) \bar{\eth }^{2}\Lambda \\
& =\left[ \partial _{1},\bar{\eth }\right] ^{1}\eth ^{3}\Lambda +%
\left[ \partial _{1},\eth \right] ^{1}\bar{\eth }^{3}\Lambda +\left[
\overline{\eth },\partial _{-}\right] ^{1}\bar{\eth }^{2}\Lambda +\left[
\eth ,\partial _{+}\right] ^{1}\eth ^{2}\overline{\Lambda }- \\
& -\left[ \left( \partial _{1}\overline{\Lambda }\partial _{-}\Lambda
+\partial _{1}\eth \overline{\Lambda }\partial _{1}\Lambda \right) \partial
_{+}\bar{\eth }^{2}\Lambda +\left( \partial _{1}\overline{\Lambda }\partial
_{-}\overline{\eth }\Lambda +\partial _{1}\eth \overline{\Lambda }\partial
_{1}\overline{\eth }\Lambda \right) \partial _{1}\bar{\eth }^{2}\Lambda +cc%
\right] .
\end{align}

From the equations (\ref{24}), (\ref{28}), (\ref{32}),
and (\ref{A4}), we can easily determine the decay in $R$ of the three last
terms in (\ref{mii}). This is not the case for $\partial_{\overline{\omega}}%
\bar{\eth }^{3}\Lambda+\partial_{\omega}\eth ^{3}\overline{\Lambda}$ and an
auxiliary expression must be found. It is possible to relate these two terms
with the free data\ by a manipulation in the two metricity conditions.
First, we equate the spin weight of the metricity conditions performing

\bigskip$3\partial_{\overline{\omega}}(\ref{a}):\qquad\qquad$%
\begin{equation*}
3\partial_{R}\partial_{\overline{\omega}}\overline{\eth}\Lambda+3\partial _{%
\overline{\omega}}\partial_{\omega}\Lambda=3\partial_{\overline{\omega}}M_{I}
\end{equation*}

$\partial_{\omega}(\ref{12}):$%
\begin{equation*}
\partial_{R}\partial_{\omega}\eth \Lambda-3\partial_{\overline{\omega}%
}\partial_{\omega}\Lambda=\partial_{\omega}M_{II}
\end{equation*}

adding these two, we get%
\begin{equation*}
\partial_{R}\left( 3\partial_{\overline{\omega}}\overline{\eth }%
\Lambda+\partial_{\omega}\eth \Lambda\right) =3\partial_{\overline{\omega}%
}M_{I}+\partial_{\omega}M_{II}
\end{equation*}

and integrating over the variable $R$,%
\begin{equation}
3\partial_{\overline{\omega}}\overline{\eth }\Lambda+\partial_{\omega }\eth
\Lambda=4\overset{\cdot}{\sigma}_{B}+\int_{\infty}^{R}\left( 3\partial_{%
\overline{\omega}}M_{I}+\partial_{\omega}M_{II}\right) dR^{\prime }.
\label{36}
\end{equation}

We see that Eq. (\ref{36}) is a s.w.2 expression and applying $\bar{\eth}%
^{2} $ we turn it into a s.w.0 one. If we carry out the corresponding
permutations and the substitution of $\bar{\eth}^{2}\Lambda=\eth^{2}%
\overline{\Lambda}$, Eq. (\ref{36}) can be reexpressed in the following
manner:

\begin{equation}
3\partial_{\overline{\omega}}\overline{\eth}^{3}\Lambda+\partial_{\omega}%
\eth^{3}\overline{\Lambda}-6\partial_{u}\bar{\eth}^{2}\Lambda+12\partial _{R}%
\bar{\eth}^{2}\Lambda=4\overline{\eth}^{2}\overset{\cdot}{\sigma}_{B}+N
\label{37}
\end{equation}
where $N=-\overline{\eth}^{2}\int_{\infty}^{R}\left( 3\partial_{\overline {%
\omega}}M_{I}+\partial_{\omega}M_{II}^{{}}\right) dR^{\prime}-3\left[
\partial_{u},\bar{\eth}\right] ^{1}\bar{\eth}\Lambda+\left[ \partial
_{\omega},\bar{\eth}\right] ^{1}\eth\bar{\eth}\Lambda+\bar{\eth}\left[
\partial_{\omega},\bar{\eth}\right] ^{1}\eth\Lambda+3\bar{\eth}\left[
\partial_{\overline{\omega}},\bar{\eth}\right] ^{1}\bar{\eth}\Lambda+3\left[
\partial_{\overline{\omega}},\overline{\eth}\right] ^{1}\bar{\eth}%
^{2}\Lambda+6\left[ \partial_{R},\overline{\eth}\right] ^{1}\bar{\eth}%
\Lambda+4\left[ \partial_{\omega},\bar{\eth}\right] ^{1}\Lambda$.

Adding to (\ref{37}) its complex conjugate expression we obtain,%
\begin{equation*}
\partial_{\overline{\omega}}\overline{\eth}^{3}\Lambda+\partial_{\omega}%
\eth^{3}\overline{\Lambda}-3\partial_{u}\bar{\eth}^{2}\Lambda-6\partial _{R}%
\bar{\eth}^{2}\Lambda=\overline{\eth}^{2}\overset{\cdot}{\sigma}_{B}+\eth^{2}%
\overset{\cdot}{\overline{\sigma}}_{B}+N_{Re}
\end{equation*}
with $N_{Re}=\frac{1}{4}\left( N+\overline{N}\right) $ representing
second and upper order terms in $\Lambda$. Inserting this last expression in
(\ref{mii}) the following result is obtained%
\begin{equation*}
\partial_{u}\bar{\eth}^{2}\Lambda+\partial_{R}\bar{\eth}^{2}\Lambda+\frac {1%
}{2}\bar{\eth}\eth\partial_{R}\bar{\eth}^{2}\Lambda=\overline{\eth}^{2}%
\overset{\cdot}{\sigma}_{B}+\eth^{2}\overset{\cdot}{\overline{\sigma}}%
_{B}+N_{Re}-\frac{1}{2}M_{IIRe}
\end{equation*}
After a carefully study of the peeling of every term in this last expression
we write%
\begin{equation*}
\left( \partial_{u}\bar{\eth}^{2}\Lambda\right) _{R\rightarrow\infty }=%
\overline{\eth}^{2}\overset{\cdot}{\sigma}_{B}+\eth^{2}\overset{\cdot }{%
\overline{\sigma}}_{B}+\overset{\cdot}{\sigma}_{B}\overset{\cdot}{\overline{%
\sigma}}_{B}
\end{equation*}
which correspond to equation (\ref{limit}) in the main text.

\end{document}